\documentclass[a4paper,11pt]{article}
\usepackage[no-natbib-sort]{pos}
\usepackage{aas_macros}
\usepackage{hyperref}
\usepackage{gensymb}

\title{An Archival Search for Very-High-Energy Counterparts to Sub-Threshold Neutron-Star Merger Candidates}
 \ShortTitle{Archival Search for VHE Counterparts to Sub-Threshold BNS Merger Candidates}

\author*[a]{Colin~Adams}
\author[]{for the VERITAS Collaboration (a complete list of authors can be found at the end of the proceedings)}
\author[b]{Imre~Bartos}
\author[a]{K.~Rainer~Corley}
\author[a]{Szabolcs~M\'arka}
\author[c]{Zsuzsanna~M\'arka}
\author[a]{Do\u{g}a~Veske}

\affiliation[a]{Physics Department, Columbia University, New York, NY 10027, USA}
\affiliation[b]{Department of Physics, University of Florida, Gainesville, FL 32611-8440, USA}
\affiliation[c]{Columbia Astrophysics Laboratory, Columbia University, New York, NY 10027, USA}

\emailAdd{ca2762@columbia.edu}

\abstract{The recent discovery of electromagnetic signals in coincidence with gravitational waves from neutron-star mergers has solidified the importance of multimessenger campaigns for studying the most energetic astrophysical events. Pioneering multimessenger observatories, such as the LIGO/Virgo gravitational wave detectors and the IceCube neutrino observatory, record many candidate signals that fall short of the detection significance threshold. These sub-threshold event candidates are promising targets for multimessenger studies, as the information provided by these candidates may, when combined with time-coincident gamma-ray observations, lead to significant detections. In this contribution, I describe our use of sub-threshold binary neutron star merger candidates identified in Advanced LIGO’s first observing run (O1) to search for transient events in very-high-energy gamma rays using archival observations from the VERITAS imaging atmospheric Cherenkov telescope array. I describe the promise of this technique for future joint sub-threshold searches.}

\FullConference{37$^{\rm{th}}$ International Cosmic Ray Conference (ICRC 2021)\\
		July 12th -- 23rd, 2021\\
		Online -- Berlin, Germany}

\begin{document}
\maketitle

\section{Introduction}\label{sec:intro}
On August 17, 2017, a binary neutron star (BNS) merger event GW170817~\citep{abbott2017gravitational} was detected for the first time in gravitational waves by the Laser Interferometer Gravitational-Wave Observatory (LIGO) and Virgo scientific collaboration (LVC). Approximately 1.7~s later, the \textit{Fermi} Gamma-Ray Burst Monitor (\textit{Fermi}-GBM) recorded a short gamma-ray burst (GRB) 170817A~\citep{abbott2017multi} in spatial coincidence with the BNS merger. The ensuing multimessenger campaign on the source revealed a kilonova in host galaxy NGC~4993 with a multi-wavelength afterglow. Two years later, another BNS merger, GW190425, was detected in the first half of LIGO/Virgo's third observing run~(O3a)~\citep{2020ApJabbottGW190425}. This merger was followed-up extensively across the electromagnetic (EM) spectrum, but notably, no EM counterpart was detected. Around the time of the merger, \textit{Fermi}-GBM observed 55.6\% of the probability region and had no onboard trigger, and found no counterpart candidate within their automated blind search or their targeted coherent search~\citep{2019GCN.24185.190425fermi}. The \textit{Fermi}-GBM team speculates that, in the case of a detectable relativistic jet, the signal likely originated from the region of the LVC localization behind the Earth for the \textit{Fermi} telescope at that time.

The current generation of imaging atmospheric Cherenkov telescopes (IACTs) have followed up on both sources, as well as a number of other LIGO triggers, with none reporting very-high energy (VHE; $> 100$ GeV) emission~\cite{santander2019recent, abdalla2017tev, ashkar2019searches, seglar2019searches}. However, all these observations took place at least an hour after the time of the purported GW event. As IACTs are sensitive primarily in the VHE band, at this point, there remains no experimental evidence of VHE emission from such events. Despite this, results from H.E.S.S. and MAGIC observations of GRBs show promise for the ability of IACTs to make a detection of a short GRB stemming from a BNS merger. Since 2016, four long GRBs have been detected by the two observatories~\citep{abdalla2019very, magic2019teraelectronvolt, 2020ATel14275.201216c, abdalla2021revealing}, and a $\sim3~\sigma$ hint of a short GRB (160821B) has been reported by MAGIC~\citep{acciari2021magic}, which began observing $\sim24$~s after the onset of the burst. As BNS mergers are expected to be a source class of short GRBs~\citep{wu2019gw170817}, this result is particularly encouraging for the future of such work.

With just two BNS merger GW detections to date, the sample size for triggered observations is quite small. However, many BNS merger candidates do not reach the threshold for detection, and are still able to be flagged in the various LVC detection pipelines. In Advanced LIGO's first observing run (O1), 103 of these sub-threshold BNS merger candidates, with a false-alarm-rate of less than one per day, were identified and publicly released~\citep{magee2019sub}. These sub-threshold candidates provide a unique opportunity to probe the BNS merger - EM emission connection at the very-high energies by using them to identify archival IACT data with coincident observations. Though these investigations come with the caveat of the candidates' sub-threshold status, they also offer a wealth of new testing ground, which can only be expected to expand in scope as relevant IACTs and GW observatories improve (subsequent LVC observing runs), emerge (the Cherenkov Telescope Array (CTA)), or begin to participate (H.E.S.S. and MAGIC).

\section{Sub-threshold Method with Archival VERITAS Data}\label{sec:subthreshvts}
We first explored this prospect in the context of archival data from the VERITAS VHE gamma-ray telescope array~\citep{adams2021subthresh}, located at the Fred Lawrence Whipple Observatory in southern Arizona, USA~\citep{holder2006first}. Given its co-location with the LIGO Hanford and Livingston sites, VERITAS is primed for multimessenger studies in GWs, as the LVC detector network is most sensitive roughly above and below these two sites~\citep{bartos2019gravitational}.

\begin{figure}[h!]
    \centering
    \includegraphics[width=\columnwidth]{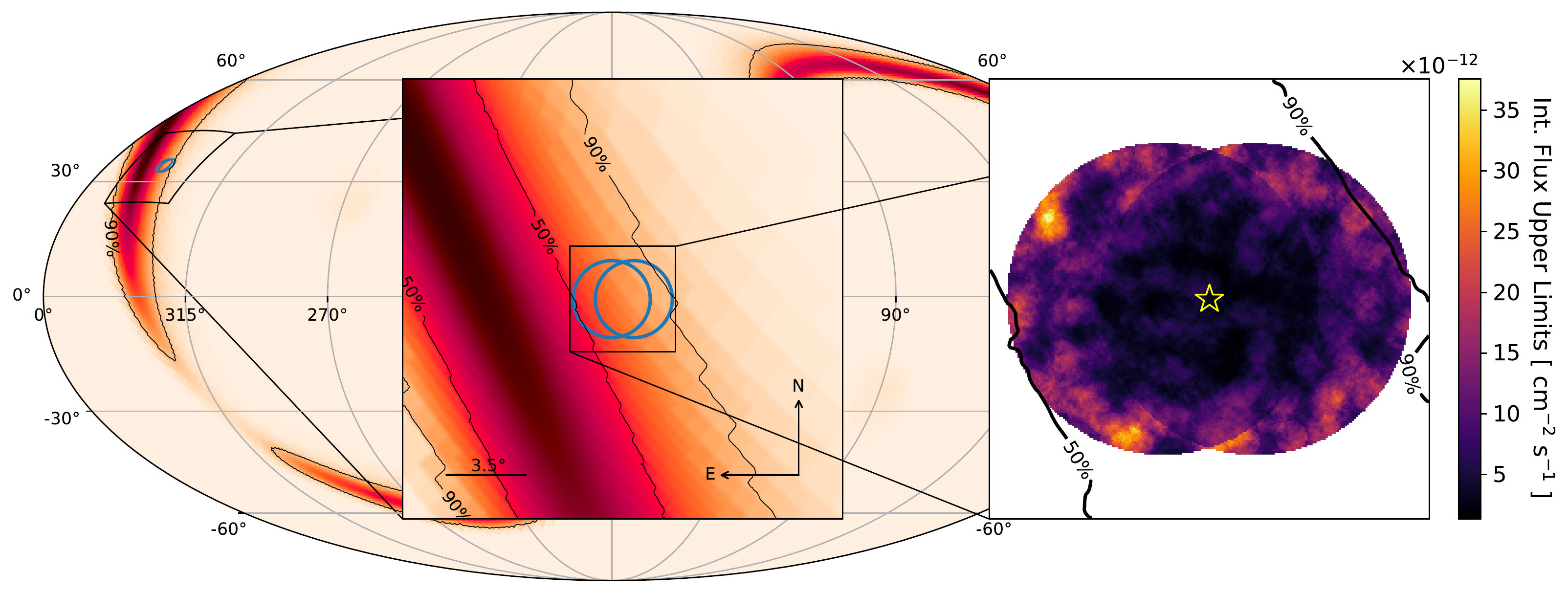}
    \caption{The localization probability map of LIGO BNS merger candidate 2015-12-04T01\_53\_39.144780 presented in equatorial coordinates. Two VERITAS observations of SN 2014c identified by the algorithm are shown to overlap spatially with the 90\% credible region of the candidate. At the right, we show the 99\% confidence bounded upper limits on the integral flux from those observations. Adapted from \cite{adams2021subthresh}.}
    \label{fig:sn2014c_ex} 
\end{figure}

This study consisted of building an algorithm to identify archival VERITAS observations (at nominal operating voltage and with zenith $< 55\degree$) that were co-located with the 90\% credible region of the sub-threshold candidate and contemporaneous within $-10 \leq t_{0} \leq 10^4$~seconds, where $t_{0}$ is the coalescence time of the event candidate. The algorithm found 11 sets of archival VERITAS observations in spatial and temporal coincidence with 7 sub-threshold BNS merger candidates. We searched each set of observations for significant excesses, and considering the large sky region covered of $\sim 98$ square degrees, also placed upper limits on each sky region covered. No significant excesses were found with a pre-trial significance above $5~\sigma$, and estimates of the incurred trials in searching such a large portion of the sky indicate that all post-trials significances are consistent with the background hypothesis. An example of the sequence in this analysis is demonstrated by \autoref{fig:sn2014c_ex}. In this figure, we show the LIGO sub-threshold candidate 2015-12-04T01\_53\_39.144780 in the background. In the foreground, we show the VERITAS observational overlaps with the localization, as well as the bounded 99\% confidence interval upper limits on the integral flux above the analysis energy threshold in the sky region covered by those observations.

Considering the astrophysical probability of the 7 coincident VERITAS observations, and VERITAS's coverage of each, we estimate the probability that at least one truly astrophysical merger was observed by VERITAS with exact spatial coincidence in the search time window to be 0.04\%.

\section{Future prospects}\label{sec:future}
\begin{figure}[h!]
    \centering
    \includegraphics[width=\columnwidth]{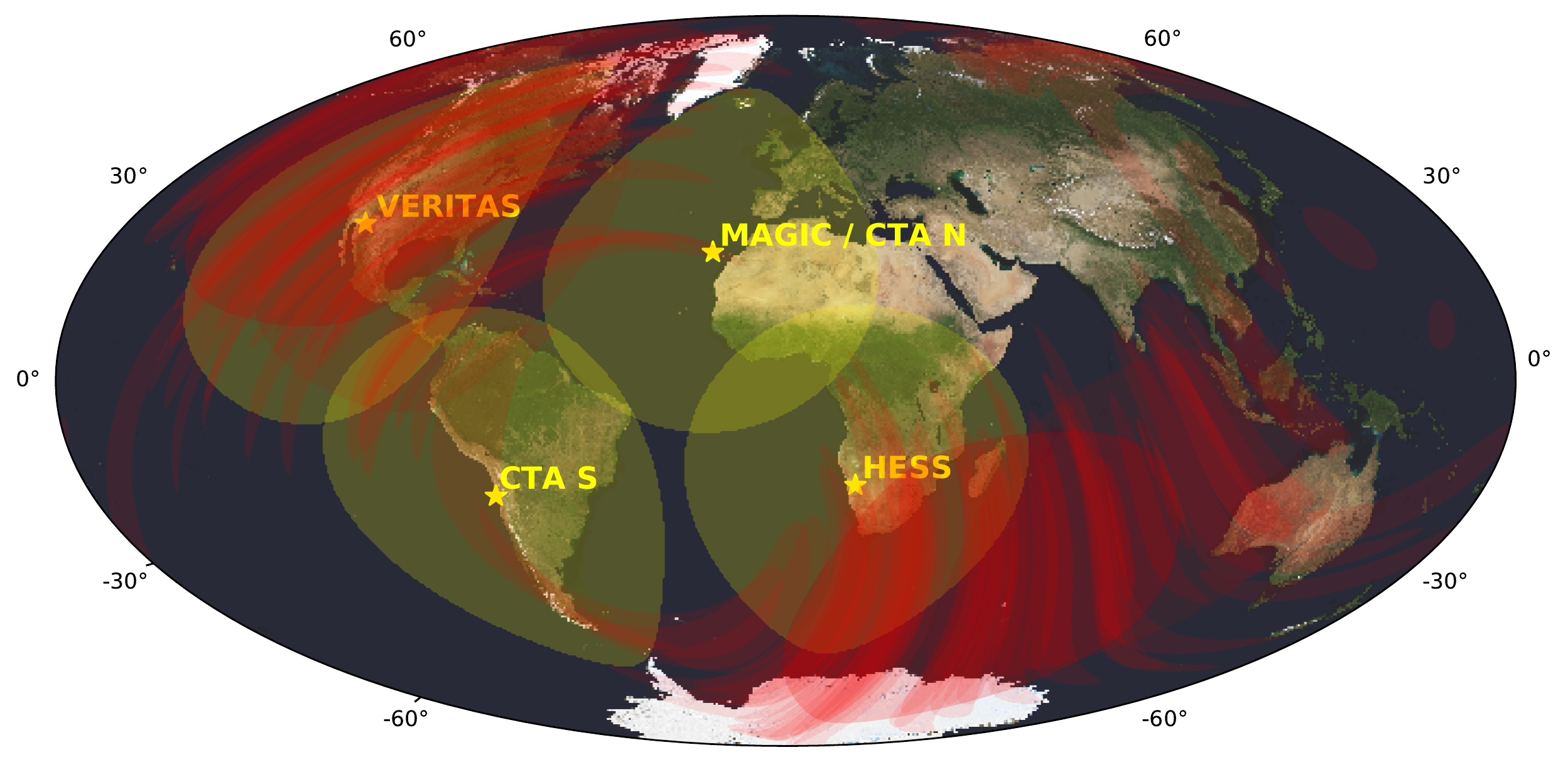}
    \caption{The 50\% credible regions of the 103 sub-threshold candidates from \citep{magee2019sub} are shown in red and are presented in geographic coordinates. The sky coverage of the current (VERITAS, MAGIC, H.E.S.S.) and future (CTA North\protect\footnotemark\ and CTA South) IACT arrays at any given time, assuming a maximum zenith of 40$\degree$, are shown in yellow.}
    \label{fig:sky_cov}
\end{figure}

As discussed at the start of \autoref{sec:subthreshvts}, VERITAS is particularly well positioned to serendipitously observe sub-threshold BNS merger candidates in archival data. However, other current-generation IACTs, particularly H.E.S.S., are also well-suited to contribute given their complementary sky coverage. This can be seen in \autoref{fig:sky_cov}, which shows the geographic locations of the 50\% credible regions for all O1 sub-threshold candidates on top of the < 40$\degree$ zenith angle coverage of all current and future-generation IACTs. The choice of the 50\% credible region in this figure was made purely for ease in visualization: if the 90\% credible regions were plotted, they would cover nearly the entire map. Notably, all present and future sky coverage regions (yellow shaded areas in the figure) overlap with multiple 90\% credible regions for BNS merger candidates. It is reasonable to expect that a number of these candidates were coincidentally observed by other current-generation instruments under the same conditions of the algorithm described above. The geographic bias in GW detections, though non-negligible, does not preclude the participation of any IACT in such a program in the past, present, or future.

\footnotetext{CTA N is co-located with MAGIC.}

Beyond this, there are also valuable real time applications to considering sub-threshold events. In the context of transmission of alerts under a Memorandum of Understanding, it would be marginally disruptive to adjust nightly observing schedules in the case of a sub-threshold candidate alert to favor higher probability regions closer to purported coalescence times. In addition, there are now 2.5 observing seasons worth of LIGO data for which sub-threshold candidates remain to be published. If the O1 performance is scaled to these new observing seasons, considering improvements in instrumentation and changing duty cycles, VERITAS could expect $\sim 70$ additional coincident observations with candidates. 

Further, such studies will benefit greatly from the planned improvements to IACT observations in the forthcoming era of gamma-ray astronomy brought by the Cherenkov Telescope Array (CTA). Not only will CTA offer a factor of $\sim 5$ improvement in field of view (FoV) area improvement over VERITAS, but also the possibility of observations with a divergent pointing strategy that could see a factor of 16 improvement \citep{gerard2016divergent}.

Both in the present and in the future, this method is suited well in the pursuit of extracting more information from archival data. Future data releases offer a useful testing ground for potential signal enhancement of sub-threshold BNS merger candidates, as well as the prospect for the detection of VHE emission from such an event. Although this study only had a 0.04\% probability of observing a true source at VHEs contemporaneously, as statistics improve with broader participation, there may also be opportunities to constrain emission at VHEs.

\acknowledgments{
This research is supported by grants from the U.S. Department of Energy Office of Science, the U.S. National Science Foundation and the Smithsonian Institution, by NSERC in Canada, and by the Helmholtz Association in Germany. This research used resources provided by the Open Science Grid, which is supported by the National Science Foundation and the U.S. Department of Energy's Office of Science, and resources of the National Energy Research Scientific Computing Center (NERSC), a U.S. Department of Energy Office of Science User Facility operated under Contract No. DE-AC02-05CH11231. We acknowledge the excellent work of the technical support staff at the Fred Lawrence Whipple Observatory and at the collaborating institutions in the construction and operation of the instrument.
}

\bibliographystyle{JHEP}
\bibliography{biball}



\clearpage
\section*{Full Authors List: VERITAS Collaboration}

\scriptsize
\noindent
C.~B.~Adams$^{1}$,
A.~Archer$^{2}$,
W.~Benbow$^{3}$,
A.~Brill$^{1}$,
J.~H.~Buckley$^{4}$,
M.~Capasso$^{5}$,
J.~L.~Christiansen$^{6}$,
A.~J.~Chromey$^{7}$, 
M.~Errando$^{4}$,
A.~Falcone$^{8}$,
K.~A.~Farrell$^{9}$,
Q.~Feng$^{5}$,
G.~M.~Foote$^{10}$,
L.~Fortson$^{11}$,
A.~Furniss$^{12}$,
A.~Gent$^{13}$,
G.~H.~Gillanders$^{14}$,
C.~Giuri$^{15}$,
O.~Gueta$^{15}$,
D.~Hanna$^{16}$,
O.~Hervet$^{17}$,
J.~Holder$^{10}$,
B.~Hona$^{18}$,
T.~B.~Humensky$^{1}$,
W.~Jin$^{19}$,
P.~Kaaret$^{20}$,
M.~Kertzman$^{2}$,
T.~K.~Kleiner$^{15}$,
S.~Kumar$^{16}$,
M.~J.~Lang$^{14}$,
M.~Lundy$^{16}$,
G.~Maier$^{15}$,
C.~E~McGrath$^{9}$,
P.~Moriarty$^{14}$,
R.~Mukherjee$^{5}$,
D.~Nieto$^{21}$,
M.~Nievas-Rosillo$^{15}$,
S.~O'Brien$^{16}$,
R.~A.~Ong$^{22}$,
A.~N.~Otte$^{13}$,
S.~R. Patel$^{15}$,
S.~Patel$^{20}$,
K.~Pfrang$^{15}$,
M.~Pohl$^{23,15}$,
R.~R.~Prado$^{15}$,
E.~Pueschel$^{15}$,
J.~Quinn$^{9}$,
K.~Ragan$^{16}$,
P.~T.~Reynolds$^{24}$,
D.~Ribeiro$^{1}$,
E.~Roache$^{3}$,
J.~L.~Ryan$^{22}$,
I.~Sadeh$^{15}$,
M.~Santander$^{19}$,
G.~H.~Sembroski$^{25}$,
R.~Shang$^{22}$,
D.~Tak$^{15}$,
V.~V.~Vassiliev$^{22}$,
A.~Weinstein$^{7}$,
D.~A.~Williams$^{17}$,
and 
T.~J.~Williamson$^{10}$\\
\noindent
$^{1}${Physics Department, Columbia University, New York, NY 10027, USA}
$^{2}${Department of Physics and Astronomy, DePauw University, Greencastle, IN 46135-0037, USA}
$^{3}${Center for Astrophysics $|$ Harvard \& Smithsonian, Cambridge, MA 02138, USA}
$^{4}${Department of Physics, Washington University, St. Louis, MO 63130, USA}
$^{5}${Department of Physics and Astronomy, Barnard College, Columbia University, NY 10027, USA}
$^{6}${Physics Department, California Polytechnic State University, San Luis Obispo, CA 94307, USA} 
$^{7}${Department of Physics and Astronomy, Iowa State University, Ames, IA 50011, USA}
$^{8}${Department of Astronomy and Astrophysics, 525 Davey Lab, Pennsylvania State University, University Park, PA 16802, USA}
$^{9}${School of Physics, University College Dublin, Belfield, Dublin 4, Ireland}
$^{10}${Department of Physics and Astronomy and the Bartol Research Institute, University of Delaware, Newark, DE 19716, USA}
$^{11}${School of Physics and Astronomy, University of Minnesota, Minneapolis, MN 55455, USA}
$^{12}${Department of Physics, California State University - East Bay, Hayward, CA 94542, USA}
$^{13}${School of Physics and Center for Relativistic Astrophysics, Georgia Institute of Technology, 837 State Street NW, Atlanta, GA 30332-0430}
$^{14}${School of Physics, National University of Ireland Galway, University Road, Galway, Ireland}
$^{15}${DESY, Platanenallee 6, 15738 Zeuthen, Germany}
$^{16}${Physics Department, McGill University, Montreal, QC H3A 2T8, Canada}
$^{17}${Santa Cruz Institute for Particle Physics and Department of Physics, University of California, Santa Cruz, CA 95064, USA}
$^{18}${Department of Physics and Astronomy, University of Utah, Salt Lake City, UT 84112, USA}
$^{19}${Department of Physics and Astronomy, University of Alabama, Tuscaloosa, AL 35487, USA}
$^{20}${Department of Physics and Astronomy, University of Iowa, Van Allen Hall, Iowa City, IA 52242, USA}
$^{21}${Institute of Particle and Cosmos Physics, Universidad Complutense de Madrid, 28040 Madrid, Spain}
$^{22}${Department of Physics and Astronomy, University of California, Los Angeles, CA 90095, USA}
$^{23}${Institute of Physics and Astronomy, University of Potsdam, 14476 Potsdam-Golm, Germany}
$^{24}${Department of Physical Sciences, Munster Technological University, Bishopstown, Cork, T12 P928, Ireland}
$^{25}${Department of Physics and Astronomy, Purdue University, West Lafayette, IN 47907, USA}

%
%
%

\end{document}